\documentclass{aa}
\usepackage{graphics}
\def\kms{\hbox{km\,s$^{-1}$}}
\def\mag{\hbox{$^{\rm m}$}}
\begin{document}

\thesaurus{06(08.09.2 $\eta$ Car; 08.05.3; 08.13.2; 09.02.1; 09.10.1)} 

\title{The nature of strings in the nebula around $\eta$\,Carinae \thanks{
Based on observations with the NASA/ESA Hubble Space Telescope, obtained
at the Space Telescope Science Institute, which is operated by the
Association of Universities for Research in Astronomy, Inc. under NASA
contract No. NAS5-26555.}}

\author {K.\ Weis \inst{1,2,}\thanks{Visiting Astronomer, Cerro Tololo
Inter-American Observatory, National Optical Astronomy Observatories,
operated by the Association of Universities for Research in Astronomy, Inc.,
under contract with the National Science Foundation.}
\and W.J.\ Duschl \inst{1,3}
\and Y.-H. Chu\inst{2,\star\star}
}
\offprints{K.\ Weis, Heidelberg, Germany - E-mail: kweis@ita.uni-heidelberg.de}
\mail{K.\ Weis, Heidelberg, Germany}

\institute{
Institut f\"ur Theoretische Astrophysik, Tiergartenstr. 15, D-69121
Heidelberg, Germany
\and
University of Illinois, Department of Astronomy, 1002 W. Green Street,
Urbana, IL 61801, USA
\and
Max-Planck-Institut f\"ur Radioastronomie, Auf dem H\"ugel 69, D-53121 Bonn,
Germany}

\date{Received 06 October 1998 / Accepted 05 July 1999}

\maketitle

\begin{abstract}

$\eta$ Carinae is one of the most extreme cases of a Luminous Blue
Variable star. A bipolar nebula of 17$^{\prime\prime}$ size
surrounds the central object. Even further out, a large amount of
filamentary material extends to a distance of 30\arcsec\ or about
0.3\,pc. In this paper we present a detailed kinematic and
morphological analysis of some outer filaments in this nebula
which we call {\it strings}. All strings are extremly long and
narrow structures. We identified 5 strings which have sizes of
0.058 to 0.177 pc in length and a width of only 0.002 pc. Using
high-resolution long-slit echelle spectroscopy it was found that
the strings follow a Hubble law with velocities increasing towards
larger distances from the star. With these unique properties, {\it
high collimation} and {\it linear increase} of the radial velocity
the strings represent a newly found phenomena in the structure and
evolution of nebulae around LBVs. Finally, we show that
morphologically similar strings can be found in the planetary
nebula NGC 6543, a possible PN-counterpart to this phenomenon.


\end{abstract}

\section{Introduction}

At a present day mass of $M \sim 120$\,M$_{\sun}$ and a luminosity of $L\sim 
10^{6.7}$\,L$_{\sun}$ (Humphreys \& Davidson 1994, Davidson \& Humphreys 1997) 
$\eta$ Carinae tops the Hertzsprung-Russell Diagram (HRD) and is certainly 
among the most massive stars observed as yet. Even if the recently re-discussed 
binary hypothesis for $\eta$ Car (Damineli 1996, Damineli et al.\ 1997, 
Davidson 1997) should turn out to apply, at least one component has to have a 
mass exceeding $\sim 60\,$M$_{\sun}$, which again puts it into the realm of the 
most massive stars. $\eta$ Car is a member of the stellar class of {\it 
Luminous Blue Variables\/} (LBVs), which start as main-sequence O stars with 
masses $M_{\rm ZAMS}$ $\ge 50$\,M$_{\sun}$; these stars evolve towards cooler 
temperatures at the end of hydrogen-core burning and may enter an unstable 
phase at an age of roughly 3\,10$^6$ years (Langer et al. 1994). This so-called 
LBV phase starts when the stars reach the Humphreys-Davidson limit (Humphreys 
\& Davidson 1979, 1994) in the HRD. Analyzing HRDs of the Galaxy and the LMC, 
Humphreys (1978, 1979) and Humphreys \& Davidson (1979) found a lack of very 
luminous red supergiants. Obviously the most massive stars do not evolve into 
red supergiants but instead their evolution is reversed towards the blue 
supergiant part in the HRD when they approach the Humphreys-Davidson limit as 
LBVs. 

One of the most prominent characteristics of the unstable LBV phase is a very
high mass loss rate (characteristically about several
$10^{-4}$\,M$_{\sun}\,{\rm yr}^{-1}$ with values even higher during giant
eruptions). Strong stellar winds and giant eruptions peel off parts of the
stellar envelope and form small circumstellar nebulae around LBVs, so-called
LBV nebulae (LBVN; Nota et al.\ 1995). In the same manner $\eta$ Car formed its
nebula in a quite dramatic way. Having been a $\sim 6$\mag\ star for a long
time (with only small changes) $\eta$ Car drastically brightened around 1843 AD 
(Herschel 1847, Innes 1903, van Genderen \& Th\'e 1984, Viotti 1995) and became 
a $\sim -$1\mag\ star for about 5 years. This giant eruption led to the 
formation of a nebula that was found only a century later. Nearly 
simultaneously Gaviola (1946, 1950) and Thackeray (1949, 1950) photographed the 
nebula for the first time. Because of its odd man-like shape Gaviola named it 
the {\it Homunculus}. Later it became clear that the Homunculus is only the 
brightest region of a larger bipolar nebula, consisting of two lobes of $\sim 
8\dots 9 $\arcsec\ diameter each. They are separated by an equatorial plane 
structure (Duschl et al.\ 1995). The high resolution Hubble Space Telescope 
(HST, Morse et al.\ 1998) images support this model, showing clearly the 
bipolarity of the Homunculus. The deepest HST pictures (200\,s in the F656 and 
F658N filters) reveal an even larger nebula consisting of a variety of 
filamentary structures like knots, arcs and strings at a distance of up to 
30\arcsec\ from $\eta$ Car. The sizes of filaments vary in a wide range from 
fractions of an arcsecond to several arcseconds. 

Analysis of the kinematics of the Homunculus and the filaments has contributed 
considerably to the understanding of the structure and nature of the nebula. 
Radial velocity (Meaburn et al.\ 1987, 1993, 1996, Hillier \& Allen 1992) and 
proper motion (Walborn 1976, Walborn et al. 1978, Walborn \& Blanco 1988, 
Currie et al.\ 1996, Smith \& Gherz 1998) measurements revealed velocities up 
to $10^3\,{\rm km}\,{\rm s}^{-1}.$ 

A comprehensive study of the full outer filamentary nebula of $\eta$ Car will
be found in Weis \& Duschl (1999, in prep.). In the present paper we 
concentrate on several very narrow, long and coherent structures which we will 
refer to as {\it strings\/}. As yet, we investigated in detail the morphology 
of the 5 most prominent strings. Moreover, we show additional kinematic 
analysis of 3 of these strings, including a full velocity coverage of the 
longest string. With the high-resolution images of the HST it was possible for 
the first time to study the LBVN of $\eta$ Car in such a detail that structures 
as narrow as the strings could be resolved and analyzed. This paper describes 
the morphology of the strings and presents their kinematic properties. We also 
discuss possible formation mechnism of the strings and address their 
uniqueness. 

\section{Observation and data reduction}

\subsection{Imaging}

\begin{figure}
\resizebox{\hsize}{!}{\includegraphics{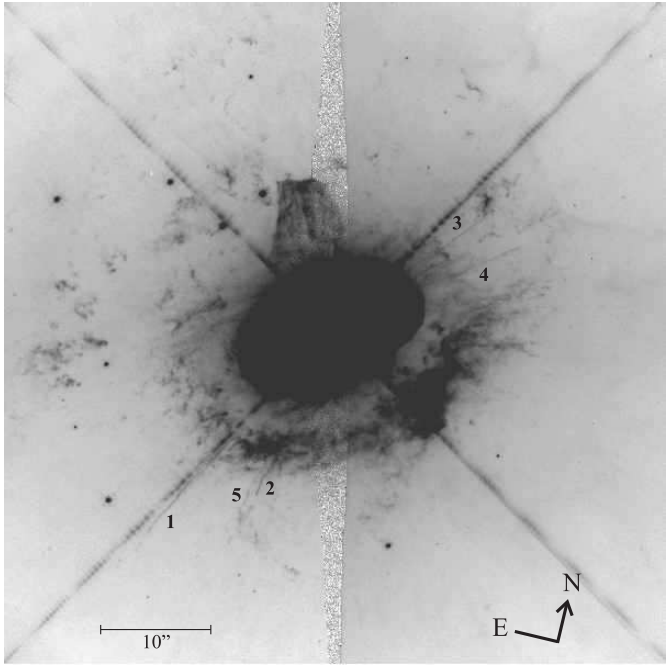}} \caption{HST picture of 
the nebula around $\eta$ Car composed of frames of different exposure times in 
the F656N filter. The field of view (FOV) is 60\arcsec$\times$60\arcsec. The 
five morpologically identified strings are indicated. North and east markers in 
the plot indicate the orientation of the image.} \label{fig:strings} 
\end{figure}

All strings were first detected in the high-resolution images of the Homunculus
nebula made with the HST, as yet the only telescope able to resolve such small
scale structures (width: 0\farcs2 - 0\farcs4 or $\sim 10^{-2}\,$pc). All HST
images we used were taken from the Canadian Astrophysics Data Center (CADC)
archive and were recalibrated using their optimal calibration data. We
retrieved all frames obtained in the F656N (H$_\alpha$) and F658N ([N\,{\sc
ii}]) filters\footnote{F656N -- program number: 5239; P.I.: J.A. Westphal;
dataset names: U2DH0101T \dots U2DH0106T; F658N -- program number: 5188; P.I.:
W.\ Sparks; dataset names: U2410501T, U2410502P \dots U2410506P}. The strings
were not visible in any other filters. In each filter the observations were
carried out with three different exposure times, 0.011, 4 and 200\,s. For
reduction, combination of the images and cosmic-ray cleaning we followed the
standard procedures recommended for WFPC2 data. Frames of the same filter were
combined weighting and scaling with the exposure time and used to correct the
bleeding by substituting the pixels with bleeding by pixels from frames with a
lower exposure time. Since most of the features seen in the nebula around
$\eta$ Carinae have expansion velocities of several $\times 10^2$\,km\,s$^{-1}$
(Meaburn et al. 1987, 1996, Hillier \& Allen 1992, Weis et al.\ in prep., and
this paper) many of the features seen in the F656N filter are contaminated by
blueshifted [N\,{\sc ii}]-emission at 6583\,\AA\, and emission seen in the
F658N filter originates in redshifted emission from the H$_\alpha$ line. Due to
the Doppler shifts, the two filters are no longer genuine H$_\alpha$ and
[N\,{\sc ii}] filters for the expanding material of $\eta$ Car. This effect is
also responsible for the wavelength-dependent length of string 1, which is
longer in the F656N than in the F658N image. While we only show images of the
F656N filter in the paper we always compare the measured sizes in both narrow
band images, keeping the Doppler shifts in mind. Figure \ref{fig:strings} shows
a 60\arcsec $\times$ 60\arcsec\ section of the final F656N WFPC2 frame. A
north-east vector indicates the celestial directions. We restrained from
rotating the images because of the loss of resolution when using the IRAF
rotation task. All images therefore have the original HST sampling of
0\farcs0455 pixel$^{-1}$ for the PC and 0\farcs0996 pixel$^{-1}$ for the WFC.

\subsection{Long-slit echelle spectroscopy\label{sect:echelle}}

To obtain kinematic information on the strings we used the echelle
spectrograph on the 4\,m telescope at the Cerro Tololo
Inter-American Observatory. We observed in the long-slit mode,
inserting a post-slit H$_\alpha$ filter (6563/75\,\AA) and
replaced the cross-disperser with a flat mirror. The
79\,l\,mm$^{-1}$ echelle grating was used, where the slit-width
was 250\,$\mu$m ($= 1\farcs 64$) resulting in an instrumental FWHM
at the H$_\alpha$ line of about 14\,km\,s$^{-1}$.

Data were recorded with the long focus red camera and the $2048 \times 2048$
Tek2K4 CCD was used. The pixel size was 0.08\,\AA\,pixel$^{-1}$ along the
dispersion, and 0$\farcs$26\,pixel$^{-1}$ on the spatial axis. Due to
vignetting, the slit length was limited to  $\sim4^\prime$.  Seeing was $\sim
2\arcsec$ during the observations and the weather was not photometric.
Thorium-Argon comparison lamp frames were taken for wavelength calibration and
geometric distortion correction. The slit positions were referenced with
respect to Th\'e et al.'s (1980) star \#66 in the Trumpler 16 cluster.

Five positions were observed covering three of the five visually identified 
strings. The spectra were taken 30\arcsec , 32\arcsec , 34\arcsec , 36\arcsec , 
and 38\arcsec\ south of our reference star (Slits 30S, 32S, 34S, 36S, and 38S, 
respectively). The slit was rotated to a position angle of PA$=132\degr$ to 
aligne it with the major axis of the Homunculus nebula and the general 
direction of the strings. Figures \ref{fig:echellograms} a-e show the spectra 
at the five positions, the strings are marked. The exposure time was 30\,s for 
slit 30S, 90\,s for slit 32S and 180\,s for the slits 34S, 36S and 38S. 

\section{Identification and morphology of the strings}

\begin{figure}
\resizebox{\hsize}{!}{\includegraphics{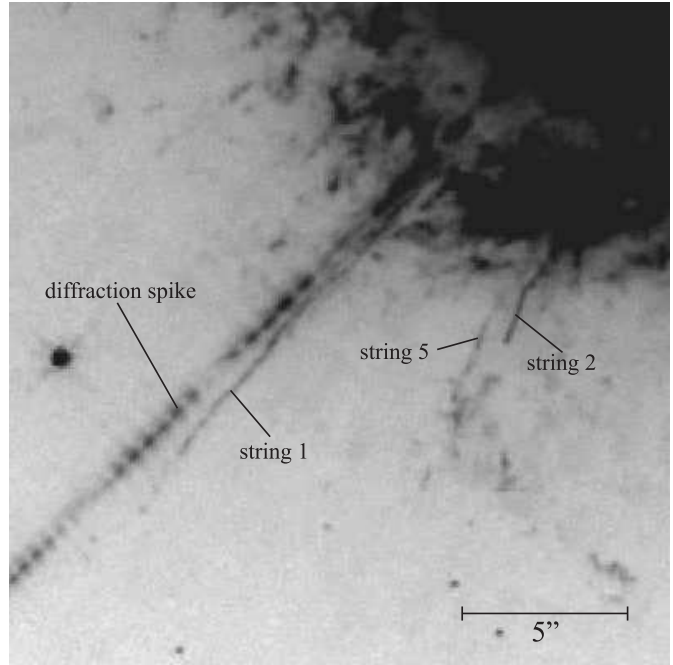}} 
\caption{Blow up of the 
HST picture of the nebula around $\eta$ Car for a more detailed presentation of 
strings 1 (long string to the left), 2 (shortest to the right) and 5 (very 
faint in the middle).} \label{fig:blow125} 
\end{figure}

\begin{figure}
\resizebox{\hsize}{!}{\includegraphics{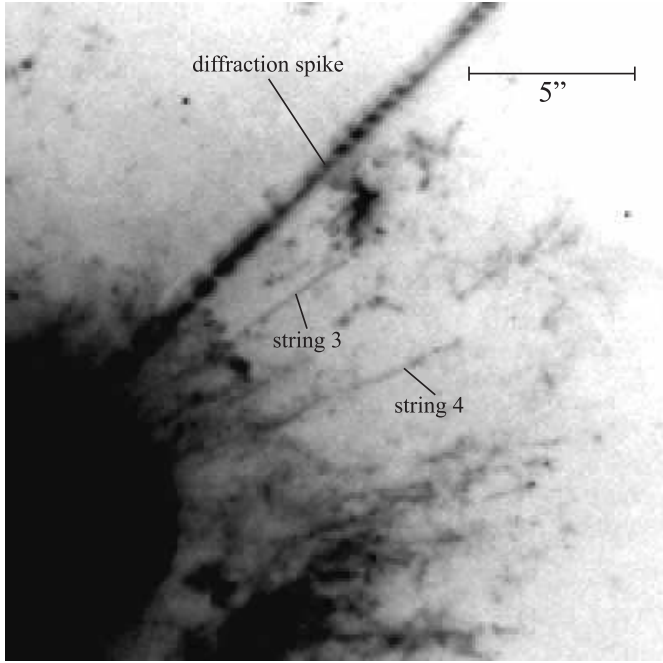}} 
\caption{Blow up of the 
HST picture of the nebula around $\eta$ Car for a more detailed view of strings 
3 (upper) and 4 (lower). } \label{fig:blow34} 
\end{figure}

A large number of filamentary structures can be identified in the deep HST 
picture of the Homunculus nebula around $\eta$ Car. Among all these 
morphologically different structures a few long, coherent and very narrow 
features are the most amazing objects we found. These features are the above 
introduced {\it strings\/}. We identified 5 such strings by visual inspection.  
On smaller length scales, morphologically similar features can be seen (Weis \& 
Duschl 1999 in prep.). However, due to confusion with background emission in 
the spectra, a detailed analysis was not possible in the work presented here. 
Of those 5 strings 3 were found in the south-eastern, and 2 in the 
north-western part of the nebula. None were detected in the other two 
quadrants. The strings are marked and named in Fig.\ \ref{fig:strings}. A 
blowup of the HST image in Fig.\ \ref{fig:blow125} gives a closer view of 
strings 1, 2 and 5 while Fig.\ \ref{fig:blow34} shows string 3 and 4 in more 
detail. 

The observed lengths of the strings range between 4\farcs0 and 15\farcs9.
Towards the center of $\eta$ Car we cannot distinguish the strings from the
overall emission of other knots and filaments. At the very end of the strings
the surface brightness decreases, and the strings might extend further at a
level below the detection limit.  Therefore, the far end of the strings is not
well determined, and all measured lengths are only lower limits.

To determine the lengths of the strings we adopt a distance of
2.3\,kpc to $\eta$ Car (Walborn 1995, Davidson \& Humphreys 1997).
However, the distance to the Carina H\,{\sc ii} region still has
not been accurately determined because of the strong and variable
reddening in this region. In the following, all sizes are measured
as a width at the base rather than FWHM.

{\it String 1\/} lies in the south-eastern part of the nebula (see
Figs.\ \ref{fig:strings} and \ref{fig:blow125}) and is the longest
of all detected. The total measured length is $15\farcs86 =
0.177\,{\rm pc} \sim 36\,500$\,AU. This length is comparable to
the combined size of the two lobes of the central bipolar nebula.

We used the PC F658N images to determine the width of string 1. In this image 
the inner part (closer to $\eta$ Car) of the string was resolved. The images 
show a width of about 5 pixels, corresponding to $0\farcs23 = 0.003\,{\rm pc} 
\sim 500\,$ AU). The resolution of the WFC is insufficient; at 2 pixels the 
strings cannot be reliably resolved, but the WFC image suggests that the width 
does not change significantly along the string. This leads to a length-to-width 
ratio of $\sim$ 70. 

If the string extends all the way to the star, an assumption supported by the
orientation of the string with respect to $\eta$ Car, the total projected
length of the string is $29\farcs0 = 0.323\,{\rm pc} \sim 66\,700$\,AU and the
length-to-width ratio would be 128. However, in the following, when not stated
explicitly otherwise, we always give the observed rather than the extrapolated
length-to-width ratio, i.e., a lower limit for this quantity.

Even though string 1 seems very straight, small, almost periodic
wiggles occur at a scale of several arcseconds as one can see in
Fig.\ \ref{fig:blow125}. At its inner end, string 1 seems either
to have a split of a length of $\sim 1\farcs5$ or to be projected
onto another string-like feature in the fore- or background. The
intensity varies along the string and decreases rapidly at its
outer end. No periodic or symmetric variations were found. Note
that String 1 corresponds to the structure called {\it jet\/} and
{\it spike\/} by Meaburn et al.\ (1987) and Meaburn et al.\
(1996), respectively.

{\it String 2\/} is shorter than String 1 and has a higher surface
brightness. It lies close to String 1 in the south-eastern part of
the nebula. String 2 is 4\farcs0 long ($= 0.044\,{\rm pc} \sim
9\,000$\,AU). The width was determined in the same way as for
string 1. A total width of 0\farcs13 ($= 0.002\,{\rm pc} \sim {\rm
nearly\ }400\,$AU) was derived leading to a length-to-width ratio
of 31. The string's far end is located 16\farcs5 ($= 0.184\,{\rm
pc} \sim 38\,000\,$AU) from the star.

The morphology of string 2 is very different from that of string
1. In contrast to the nearly straight line image of string 1,
string 2 shows a prominent kink (Fig.\ \ref{fig:blow125}) where
the direction of the string changes by 22\degr $\pm$2\degr. String
2 has a more uniform surface brightness and has a clearer end
where the surface brightness drops abruptly.

{\it String 3\/} is one of the two strings found in the
north-western part of the Homunculus nebula (Fig.\
\ref{fig:blow34}). We derive a length of 7\farcs6 ($ = 0.085\,{\rm
pc} \sim 17\,500\,$AU). The string is covered in both the WFC and
the PC images in its full length, so that the width of the string
over its entire observed length can be measured. The width is
$0\farcs19 = 0.002\,$pc ($\sim 400\,$AU) and does not change
significantly along the string. This results in a length-to-width
ratio of 42. Extrapolating the string back to the star we derive a
length of 17\farcs0 ($= 0.189\,{\rm pc} \sim 39\,000\,$AU). String
3 is the straightest of all strings and shows only one small
wiggle at its outer end. The string's surface brightness is very
homogenous, showing nearly no changes in its intensity.

{\it String 4\/} lies close to string 3 (Fig.\ \ref{fig:blow34}) and has
similar parameters. Its length amounts to 9\farcs3 ($= 0.103\,{\rm pc} \sim 
21\,400\,$AU). The width measured on the PC image is 0\farcs14 ($= 0.002\,{\rm 
pc} \sim 300\,$AU) and is approximately constant along the string. The ratio of 
length-to-width is 68. Altogether, the string extends 18\farcs7 ($= 0.208\,{\rm 
pc} \sim 43\,000\,$AU) from the star. String 4 bends and wiggles slightly but 
less frequently than string 1. 

{\it String 5\/} is much harder to identify due to its low surface brightness. 
This might also be the reason why it seems less coherent and hard to resolve. 
It might easily be mistaken as a number of individual filaments. String 5 was 
measured to be 5\farcs2 long ($= 0.058\,{\rm pc} \sim 12\,000\,$AU) and 
0\farcs14 ($= 0.002\,{\rm pc} \sim 300\,$AU) wide. This gives a length-to-width 
ratio of 38. A length of 18\farcs7 ($= 0.208\,{\rm pc} \sim 43\,000\,$AU) was 
obtained for the full distance from $\eta$ Car to the outer end of string 5. 

String 5 shows more brightness variations along its extension than the others, 
giving it a somewhat knotty appearance. The three brightest regions of the 
string can be identified in Fig.\ \ref{fig:blow125}. String 5 shows one larger 
bend, but no small scale wiggles are found. 

Note also that at their far ends from the star, strings 1 and 5 become 
brighter, in contrast to the other strings. In string 1 this brightening looks 
like two knots.

\begin{table}
\caption[]{Properties of the strings}
\begin{flushleft}
\begin{tabular}{ccccc}
\hline
 & observed \\
\raisebox{1.5ex}[-1.5ex]{string} & length &
\raisebox{1.5ex}[-1.5ex]{$v_{\rm min}$} &
\raisebox{1.5ex}[-1.5ex]{$v_{\rm max}$} &
\raisebox{1.5ex}[-1.5ex]{[N\,{\sc ii}]$\lambda$ 6583/H$_\alpha$}
\\ & [pc] & [km\,s$^{-1}$] & [km\,s$^{-1}$] & \\ \hline 1    &
0.177 & $-$522 & $-$995 & 3.3\\ 2    & 0.044 & $-$442 & $-$591 &
2.7\\ 3    & 0.095 & - & - & - \\ 4    & 0.103 & - & - & - \\ 5
& 0.058 & $-$383 & $-$565 & 3.0\\ \hline
\end {tabular}
\end{flushleft}
\label{tab:strings}
\end{table}

\section{Kinematic analysis}

\begin{figure}
\resizebox{\hsize}{!}{\includegraphics{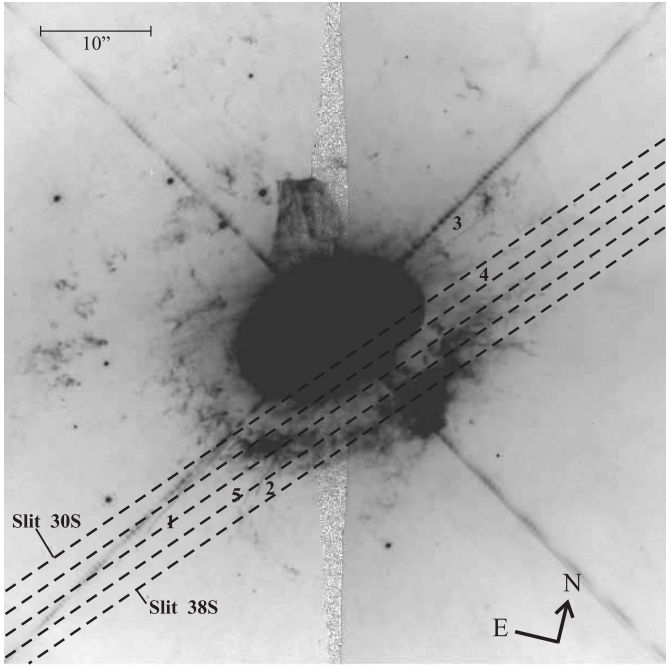}} 
\caption{Positions and 
naming convention of the slits \label{fig:slitpositions}} 
\end{figure}

\begin{figure*}
\resizebox{13.4cm}{!}{\includegraphics{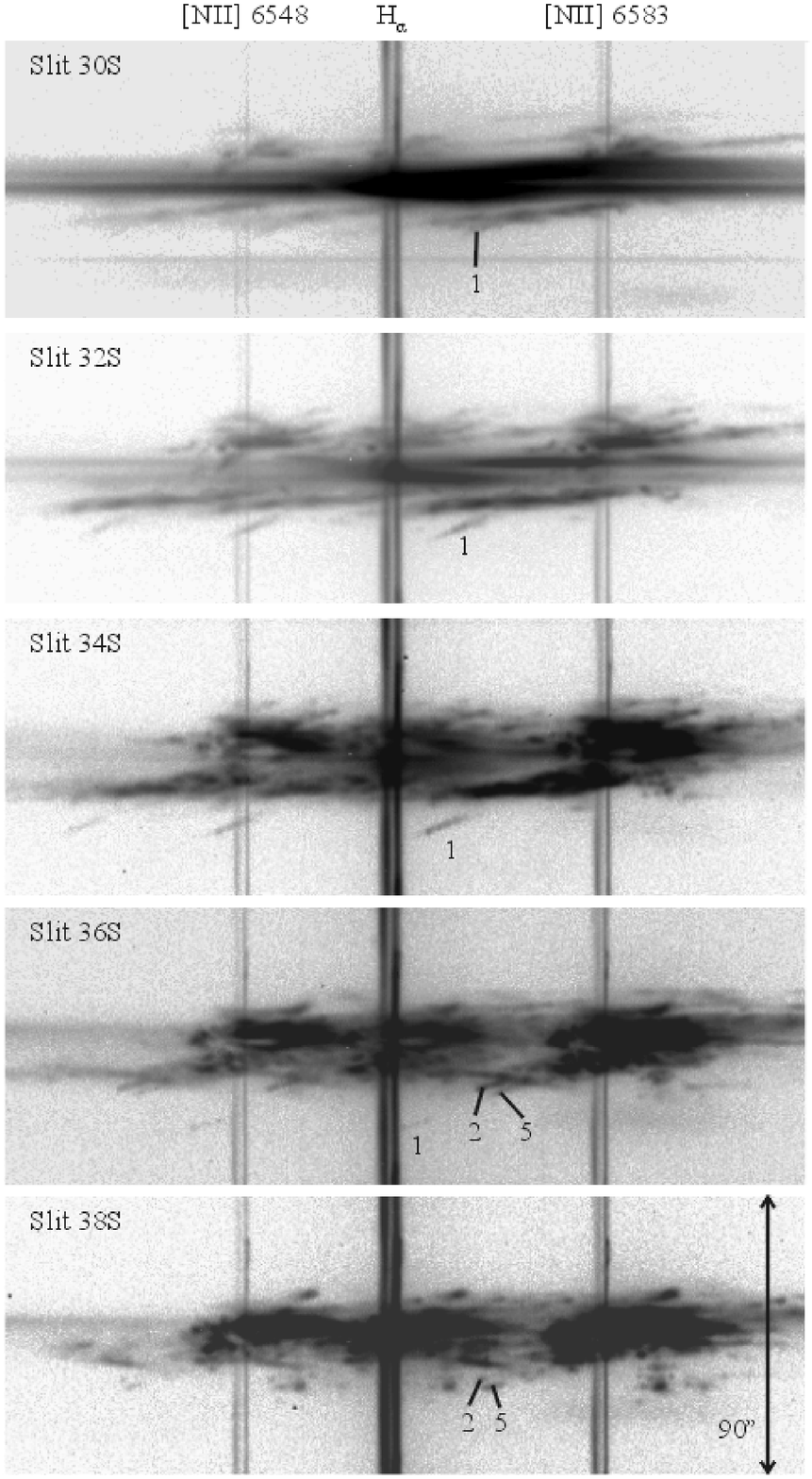}} \hfill
\parbox[b]{41mm}{
\caption{Echellograms at the five slit positions which intercept the strings. 
Continuous split lines indicate the expansion of the gas in the Keyhole nebula 
in the vicinity of $\eta$ Carinae. String 1, 2 and 5 are marked. The spectra 
were offset by 2\arcsec\ south. Due to the position angle of PA=132\degr of the 
slit the parallel offset is about 1\farcs5. } \label{fig:echellograms}} 
\end{figure*}

We obtained kinematic data for three of the five strings, using
long-slit echelle observations. String 1 was covered in Slits 30S,
32S, 34S and 36S (see Table \ref{tab:string1}). String 2 was
intercepted by Slits 36S and 38S (Table \ref{tab:string2}), string
5 was found in Slits 36S and 38S (Table \ref{tab:string5}). The
positions of the slits are shown in Fig.\ \ref{fig:slitpositions},
the corresponding echellograms in Fig.\ \ref{fig:echellograms}
a-e. The spectral axis covers 80\,\AA\, and is centred on the
H$_\alpha$ line at 6563\,\AA.

In addition to the emission from the Homunculus and its immediate surroundings, 
the spectra show prominent H$_\alpha$ and [N\,{\sc ii}]\,6548 and [N\,{\sc 
ii}]\,6583 lines which are split by about 40\,\kms\ and which are due to the 
expansion of the nebula (Deharveng \& Maucherat 1975). 

\begin{figure}
\resizebox{\hsize}{!}{\includegraphics{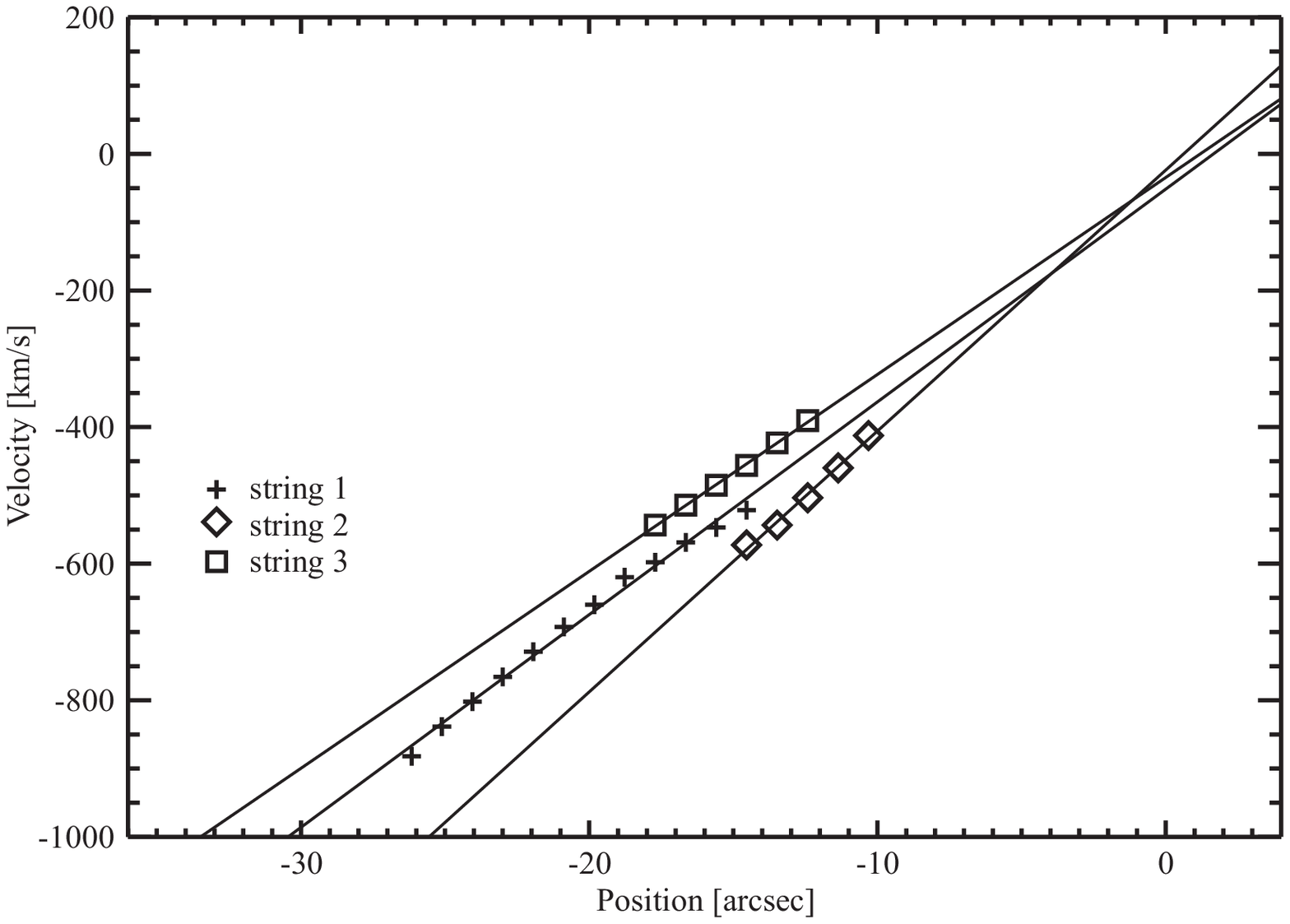}} 
\caption{The observed 
radial velocities as functions of the projected distance from the star for 
strings 1, 2, and 5.} \label{fig:posvel} 
\end{figure}

Strings 1, 2, and 5 are marked in Fig.\ \ref{fig:echellograms} at the [N\,{\sc 
ii}] 6583\,\AA\ line position. All strings show a slope in their respective 
spectra, indicating a linearly increasing velocity towards larger distances 
from the star. The radial velocities of the strings are lowest closest to 
$\eta$ Car and increase to the maximum velocity at their far ends. Figure 
\ref{fig:posvel} gives the observed radial velocities as functions of the 
projected distance from the star for the three strings. The velocity increase 
is very continuous, as seen in Fig.\ \ref{fig:echellograms}. All velocities of 
strings 1, 2, and 5 are negative, i.e., all strings in the south-eastern part 
are approaching us, in agreement with the fact that the bipolar lobes of the 
Homunculus are tilted such that the south-eastern lobe is closer to the 
observer. Within the uncertainties of our results, the extrapolation backwards 
leads to a radial velocity of 0 \kms at the position of the star. The main 
uncertainty comes from the determination of the location of the star relative 
to the positions of our spectra; we estimate this uncertainty to be about 1 -- 
2$^{\prime\prime}$. 

In the following we describe the kinematic parameters of the
individual strings:

{\it String 1\/} (Table \ref{tab:string1}): The radial velocities of string 1
range from $-$522\,km\,s$^{-1}$  at the inner part to $-$996\,km\,s$^{-1}$ at
the far end. The velocity gradient is the same in all spectra, revealing a
steady increase of $31.2\,\kms/1\arcsec = 2790\,\kms\,{\rm pc}^{-1}$. We will 
comment on this phenomenon in Sect.\ \ref{sect:disc}. 

Beside the slope, the spectra show that string 1 consists of knot-like
substructures (Fig.\ \ref{fig:echellograms}) with distiguishable velocities.
The width of the knots in velocity, typically 22 \kms , is larger than the
instrumental FWHM of 14 \kms\ implying the knots are at least marginally
resolved in velocity space. Continuous emission connects the substructures and
forms the strings. The knotty structures are most prominent in Slits 34S and
36S, they nearly disappear in Slits 32S and 30S.

In addition to the kinematics, the echelle spectra provide us with information 
about the nitrogen excitation. For string 1 we find a ratio of [N\,{\sc 
ii}]$\lambda$6583/H$_\alpha$ $\sim$ 3.3 $\pm$ 0.3 (Table \ref{tab:strings}). 
The ratio is constant along the string. It is in the same range as observed in 
other regions of the nebula around $\eta$ Car (Meaburn et al.\ 1987, 1996 
[c.f., in particular their Fig.\ 5]) 

{\it String 2\/} (Table \ref{tab:string2}): Analogous to string 1 we found a 
velocity gradient in string 2 reaching from $-$442\,\kms\ at the inner to 
$-$591\,\kms\ at the outer end of the string (Fig.\ \ref{fig:echellograms}). 
This leads to a velocity increase of $38.2\,\kms/1\arcsec = 3420\,\kms\,{\rm 
pc}^{-1}$. No knotty substructures were found. [N\,{\sc 
ii}]$\lambda$6583/H$_\alpha$ yields a value of 2.7 $\pm$ 0.3, only slightly 
different from that found for string 1. 

{\it String 5\/} (Table \ref{tab:string5}): This string shows the same 
behaviour as strings 1 and 2, namely a radial velocity increase towards the 
outer end. Starting at $-$383 \kms\ it reaches $-$565\,\kms\ at the tip. This 
translates into a gradient of $28.9\,\kms/1\arcsec = 2590\,\kms\,{\rm 
pc}^{-1}$. The change in velocity is constant. No knotty substructures were 
detected. 

Measuring the line ratio for string 5 was complicated  by a large
amount of diffuse background emission at the H$_\alpha$ line (see
Figs.\ \ref{fig:echellograms} d and e). We measured a ratio of [N
II]$\lambda$6583/H$_{\alpha}$ $\simeq$ 3.0 $\pm$ 0.3, taking into
account the background contamination.

\begin{table}
\caption[]{\small Position and velocities of string 1}
\begin{flushleft}
\begin{tabular}{crrrr}
\hline slit & \multicolumn{2}{c}{position} &
\multicolumn{1}{c}{$v_{\rm min}$} & \multicolumn{1}{c}{$v_{\rm
max}$} \\ & from [\arcsec] & to [\arcsec] & [\kms] & [\kms] \\
\hline
  Slit 30S  &  -20.1  &  -14.3 &   -667 &  -522 \\
  Slit 32S &   -25.4  &  -14.6 &  -809 &  -522 \\
  Slit 34S  &  -26.2  &  -18.0 &-853 &  -631 \\
  Slit 36S  &  -29.4  &  -23.0 & -995 &  -758 \\
\hline
\end {tabular}
\end{flushleft}
\label{tab:string1}
\end{table}

\begin{table}
\caption[]{\small Position and velocities of string 2 }
\begin{flushleft}
\begin{tabular}{crrrr}
\hline
slit & \multicolumn{2}{c}{position} &
\multicolumn{1}{c}{$v_{\rm min}$} & \multicolumn{1}{c}{$v_{\rm
max}$} \\ & from [\arcsec] & to [\arcsec] & [\kms] & [\kms] \\
\hline
 Slit 36S  &  -15.6  & -11.1 & -584 &  -442 \\
 Slit 38S  &  -16.4  &  -13.5 &  -591  & -533 \\
\hline
\end {tabular}
\end{flushleft}
\label{tab:string2}
\end{table}

\begin{table}
\caption[]{\small Position and velocities of string 5}
\begin{flushleft}
\begin{tabular}{crrrr}
\hline
slit & \multicolumn{2}{c}{position} &
\multicolumn{1}{c}{$v_{\rm min}$} & \multicolumn{1}{c}{$v_{\rm
max}$} \\ & from [\arcsec] & to [\arcsec] & [\kms] & [\kms] \\
\hline
 Slit 36S  &  -16.7   & -12.4   & -529 &  -383 \\
 Slit 38S  &  -18.0  &  -15.1 &  -565  & -456 \\
\hline
\end {tabular}
\end{flushleft}
\label{tab:string5}
\end{table}

Based on the assumption that the strings were not accelerated or decelerated 
considerably since their formation, and that they were formed together with the 
Homunculus $\sim 150\,$yrs ago, we can determine the inclination $\phi$ of the 
strings with respect to the plane of the sky\footnote{An independent 
determination of the strings' inclination and expansion age is not possible as 
we have the velocity and positional informations in directions orthogonal to 
each other only.} ($\tan \phi = v_{\rm r}\,t\,s^{-1}$, where the parameters are 
defined as follows : $v{\rm _r} := $ radial velocity $s :=$ projected length 
and $ t := $ time since eruption in 1843), and find angles of $22^\circ$ for 
string 1, $27^\circ$ for string 2, and $20^\circ$ for string 5, respectively. 
The accuracy of this determination is determined by the positional accuracy of 
our velocity measurements relative to the star; we estimate the accuracy of the 
angles to $\sim \pm 3^\circ$. This orients the strings roughly in the direction 
of the major axis of the Homunculus. 

For the strings in the north-western part of the Homunculus nebula we could not
extract unambiguous kinematic information because the spectra taken at the
positions of the strings contain too much diffuse and continuous emission. From
our measurements of other features and guided by the assumption that the
strings follow the general tilt of the Homunculus, we would be surprised if
strings 3 and 4 do not have positive radial velocities.

\section{Discussion and conclusions\label{sect:disc}}

At the end of core hydrogen burning the most massive stars enter a short ($\sim
25\,000\,$yrs) but violent phase of evolution when they turn into LBVs. This
phase is characterized by a high mass loss rate (up to $10^{-4}\,$M$_{\sun}$,
and even more during giant eruptions) and intermittent {\it giant eruptions\/}.
During these eruptions the star's visual luminosity increases by several
magnitudes. The mechanism causing this behaviour is still far from understood.
High mass loss and giant eruptions lead to the formation of the nebulae around
the LBVs. Garc\'{\i}a-Segura et al. (1996) and Dwarkadas and Balick (1998)
showed in a hydrodynamic model that the interaction of an older slow and a
younger fast wind in the LBV phase may give rise to the structure of the LBVN.

One finds LBVNs in very different shapes and sizes. A comprehensive compilation
of nebulae around LBVs known as yet may be found in Nota et al. (1995). Already
ground-based images revealed that most of the LBVNs have small-scale
substructures like knots and arcs. These features can be seen in detail in
recent HST images, e.g. of AG Car, HR Car. In particular the Homunculus nebula
around $\eta$ Car contains a large variety of substructures and individual
knots (e.g., Walborn 1976). Beside the knotty structure of the central bipolar
lobes (diameter of each lobe: $\sim 0.1\,$pc) numerous filaments exceed the 
size of the central nebula and can be found at distances up to 
$30^{\prime\prime}$ or $\sim 0.3\,$pc from the star (see Fig.\ 
\ref{fig:strings}). 

When analyzing these structures around $\eta$ Car, we find that the strings
show amazing and unexpected characteristics. They are {\it ex\-tre\-me\-ly
narrow\/} with {\it very high length-to-width ratios\/}, generally {\it very
straight\/}, and they follow a {\it perfectly linear velocity increase\/}
towards increasing distances from the star. Back-extrapolation of this linear 
dependency to the position of the star is consistent with a vanishing radial 
velocity there, i.e., we have a projected distance-radial velocity law of the 
{\it Hubble type\/}. 

\subsection{The nature of the strings}

The physical nature of the strings is far from understood yet. They may or may 
not be a single physical entity. One may think of a coherent structure, similar 
to a water jet, for instance. However, one may equally well envisage a train of 
many individual knots or bullets following the same path. One also cannot rule 
out the possibility that they are trails or a wakes following an object at the 
strings' far ends, or even projection effects of the walls of, for instance, 
much wider funnels. 

Currently, one can only speculate about the reasons for
\begin{itemize}
\item the {\it straightness\/} of the strings: This may point to an origin of
the strings at a much smaller distance from the star than their current
location, for instance at or close to the star's surface. Then even Keplerian
azimuthal velocity and the accompanying angular momentum would result in a
negligible azimuthal velocity at their current location as long as they do not
gain considerable amounts of angular momentum during their evolution, which is
certainly a reasonable assumption. Slight deviations from the straight 
expansion may be due to deflections caused by interaction with the local 
ambient medium; 
\item the {\it narrowness and collimation\/} of the strings: Most likely, it is
due to their highly supersonic expansion velocities. If -- at best -- the
lateral expansion is at thermal speeds, then this corresponds to "opening
angles" of the strings of $1^\circ$ or less, i.e., far less than can be
detected in the available body of imaging data. This seems to indicate that the
narrowness itself was defined in the origin of the strings.
\item the {\it Hubble type velocity law\/} of the strings: While a geometric
projection effect is extremely unlikely, one may think of several possible 
other mechanisms, or combinations of them. 
\begin{itemize}
\item {\it Stellar winds\/}: For stellar winds, for a certain radial range in
the vicinity of the star at several, but not many stellar radii, a linear
velocity profile is a good approximation. In our case, however, the geometrical
dimension of the strings is very much larger than the stellar radius, making
such an interpretation again very unlikely. However, if the working surface of
the acceleration process were much larger than the star itself and no longer
small compared to the radial dimensions of the string, it could give rise to
such a linear profile.
\item {\it Stellar explosions\/}: In stellar explosions -- rather than winds --
a linear velocity profile is a good approximation for the larger radii, close
to the head of the explosion (Tscharnuter \& Winkler 1979). Given the highly
supersonic velocities of the strings, such a stellar explosion may be a viable
model for the velocity evolution of the strings.
\item {\it Primordial velocity spectrum\/}: At their formation the strings were
ejected within a time scale that is short compared to the time scale of the
evolution since then, and this ejection happened with a certain distribution of
velocities.
\end{itemize}
\end{itemize}
In any case, it is noteworthy that such a linear expansion law is also observed
for the proper motion of the main structures of the Homunculus (Currie et al.
1996).

In summary, one has to conclude that the linear velocity law of the strings is 
far from being really understood. However, the orientation of the strings - 
assuming a certain epoch of formation - points towards a close relation to the 
event that also created the Homunculus, as do the results of the Hubble-type 
law.

\subsection{Are the strings unique ?}

\begin{figure}
\resizebox{\hsize}{!}{\includegraphics{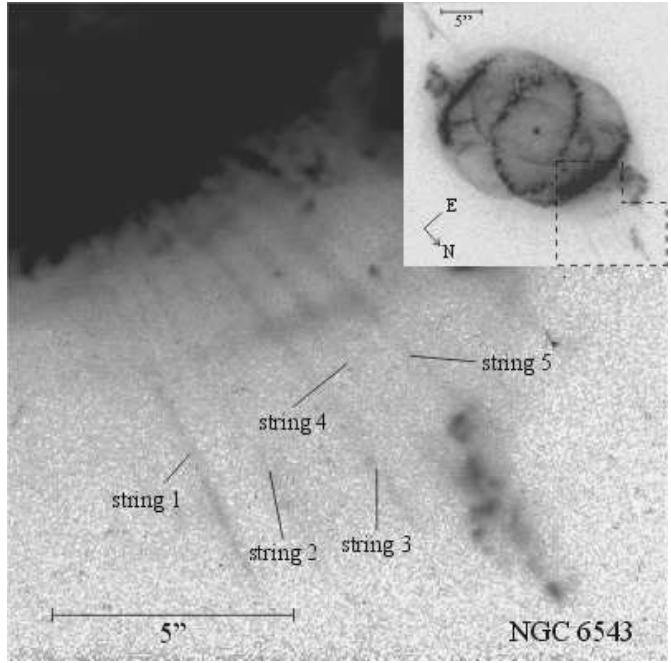}} 
\caption{The planetray 
nebula NGC 6543 observed with the HST F658N filter. In the insert the whole 
nebula with the central PN star is shown while the large image shows the 
northern section of the PN nebula with 5 string-like objects similar to what we 
identified in $\eta$ Car.} \label{fig:pn} 
\end{figure}

\begin{table}
\caption[]{Parameters of strings like structures in NGC\,6543, for the width 
the minimum and maximum values are listed} 
\begin{flushleft}
\begin{tabular}{cccc}
\hline string & length & width & length-to-width \\ & [10$^{-3}$\,pc] & 
[10$^{-3}$\,pc]  & \\ \hline 1    & 39 & 0.8-1.3 & 30-50\\ 2    & 43 & 0.5-1.3 
& 33-82\\ 3    & 39 & 0.5-1.3 & 30-74\\ 4    & 43 & 0.5-1.5 & 10-30\\ 5    & 35 
& 1.5-2.1 & 17-22\\ \hline 
\end {tabular}
\end{flushleft}
\label{tab:pn} 
\end{table}

Even though many of the LBVN show small scale structures and knots, no strings 
were found in other LBVNs. In particular for AG Car and HR Car, on HST images 
one should be able to detect narrow strings of the type seen around $\eta$ Car, 
but none were found. In the light of the ongoing discussion about morphological 
and physical similarities and relations between LBVN and planetary nebulae 
(see, e.g., Frank et al.\ 1998, Frank 1998, Currie \& Dowling 1999), it is of 
interest to look into the question whether the strings and their properties are 
unique to LBVNs. 

We re-analyzed the HST images of NGC 6543 from the CADC archive ([N\,{\sc ii}] 
F658N filter) and found at least 5 (possibly 2 more) features (Fig.\ 
\ref{fig:pn}) that resemble our strings very much (see also in Harrington \& 
Borkowki 1994, 1995). The PC again allowed us to determine the width of the 
strings between 2-8 pixel. Assuming a distance to NGC\,6543 of 1170\,pc (Castor 
et al.\ 1981) we found that the string lengths are 35 10$^{-3}$ to 43 
10$^{-3}$\,pc while they are 0.5 10$^{-3}$ to 2.1 10$^{-3}$\,pc wide (Table 
\ref{tab:pn}). Their length-to-width ratios range from 22 to 82. Comparing 
these string-like structures with the strings in $\eta$ Car we conclude that 
they show approximately the same morphology. The linear sizes are comparable. A 
comparison between these PN-strings and the ones in LBVs, which we discussed in 
the present paper, will be of utmost importance as it has the potential of 
yielding insight into the differences and into the similarities in the 
formation processes of nebular structures around evolved high (LBV) and low 
(PN) mass stars. 

\begin{acknowledgements}
The data reduction and analysis was in part carried out on a workstation 
provided by the {\it Alfried Krupp von Bohlen und Halbach Stiftung\/}. This 
support is gratefully acknowledged. We thank D.J.\ Bomans, Bochum, for many 
stimulating discussions of the subject of this paper, and the referee for 
helpful comments. Guest User, Canadian Astronomy Data Centre, operated by the 
Herzberg Institute of Astrophysics, National Research Council of Canada. 
\end{acknowledgements}

\end{document}